\documentclass[preprint,aps,a4,amsmath,amssymb,floatfix]{revtex4-1}

\usepackage{bm}
\usepackage{graphicx}
\usepackage{amsbsy}
\newcommand{\vlk}{$V_{\rm low-k}$ }
\newcommand{\vlkn}{$V_{\rm low-k}$}
\newcommand{\be}{\begin{equation}}
\newcommand{\ee}{\end{equation}}

\begin{document}

\title{Second-order quasiparticle interaction in nuclear matter \\
with chiral two-nucleon interactions\footnote{Work 
supported in part by BMBF, GSI and by the DFG cluster of excellence: Origin and 
Structure of the Universe.}}

\author{J.\ W.\ Holt, N.\ Kaiser, and W.\ Weise}
\affiliation{Physik Department, Technische Universit\"{a}t M\"{u}nchen,
    D-85747 Garching, Germany}

\begin{abstract}
We employ Landau's theory of normal Fermi liquids to study the quasiparticle 
interaction 
in nuclear matter in the vicinity of saturation density. Realistic 
low-momentum nucleon-nucleon interactions evolved from the Idaho N$^3$LO chiral 
two-body potential are used as input potentials. We derive for the first time
exact results for the central part of the quasiparticle interaction computed to second order in 
perturbation theory, from which we extract the $L=0$ and $L=1$ Landau parameters
as well as some relevant bulk equilibrium properties of nuclear matter. 
The accuracy of the intricate numerical calculations is tested with analytical
results derived for scalar-isoscalar boson exchange and (modified)
pion exchange at second order. 
The explicit dependence of the Fermi liquid parameters on the low-momentum cutoff scale
is studied, which provides important insight into the scale variation  
of phase-shift equivalent {\it two-body} potentials. This leads naturally to
explore the role that three-nucleon forces must play in the effective 
interaction between two quasiparticles.
\end{abstract}

\maketitle

\section{Introduction}

Describing the properties of infinite nuclear matter has long been an important
benchmark for realistic models of the nuclear force and the applied many-body 
methods. Recent calculations \cite{fritsch05,bogner05,siu09,hebeler11} have 
shown that the (Goldstone) linked-diagram expansion (up to at least second order) 
can provide an adequate description of the zero-temperature equation of state 
when realistic two-nucleon and three-nucleon forces are employed. 
In the present work we study nuclear
matter from the perspective of Landau's Fermi liquid theory 
\cite{landau57,migdal1,migdal2,baym}, which is a 
framework for describing excitations of strongly-interacting normal Fermi systems in terms
of weakly-interacting quasiparticles. Although the complete description of the 
interacting many-body ground state lies beyond the scope of this theory, various bulk
equilibrium and transport properties are accessible through the quasiparticle interaction.

The interaction between two quasiparticles can be obtained microscopically
within many-body perturbation theory by functionally differentiating the total 
energy density twice with 
respect to the quasiparticle distribution function. Most previous
studies using realistic nuclear forces have computed only the 
leading-order contribution to the quasiparticle interaction exactly, while 
approximately summing certain classes of diagrams to all orders 
\cite{babu73,sjoberg73,dickhoff83,backman85,holt07}. 
In particular, the summation of particle-particle ladder diagrams in the Brueckner $G$-matrix
was used to tame the strong short-distance repulsion present in most realistic nuclear 
force models, and the inclusion of the induced interaction of Babu and Brown 
\cite{babu73} (representing the exchange of virtual collective modes between 
quasiparticles) was found to be essential for achieving the stability of nuclear matter 
against isoscalar density oscillations.

To date, few works have studied systematically the order-by-order 
convergence of the quasiparticle interaction using realistic models of the nuclear force. In ref.\
\cite{kaiser06} the pion-exchange contribution to the quasiparticle interaction
in nuclear matter was obtained at one-loop order, including also the effects
of $2\pi$-exchange with intermediate $\Delta$-isobar states. In the present work we derive
general expressions for the second-order quasiparticle interaction in terms
of the partial wave matrix elements of the underlying realistic nucleon-nucleon
(NN) potential. The numerical accuracy of the second-order calculation in this
framework 
is tested with a scalar-isoscalar-exchange potential as well as a (modified)
pion-exchange interaction, both of which allow for exact analytical solutions at 
second order. We then study the Idaho N$^3$LO chiral NN interaction \cite{entem} 
and derive from this potential a set of low-momentum nucleon-nucleon interactions 
\cite{bogner02,bogner03}, which at a 
sufficiently coarse resolution scale ($\Lambda \simeq 2$\,fm$^{-1}$) provide a model-independent two-nucleon
interaction and which have better convergence properties when employed in many-body
perturbation theory \cite{bogner05,bogner10}. We extract the four components of the 
isotropic ($L=0$) quasiparticle interaction of which two are related to the  
nuclear matter incompressibility ${\cal K}$ and symmetry energy $\beta$.
The $L=1$ Fermi liquid parameters, associated with the angular dependence
of the quasiparticle interaction, are used to obtain properties 
of the quasiparticles themselves, such as their effective mass $M^*$ and the anomalous orbital 
$g$-factor. Our present treatment focuses on the role of two-nucleon interactions.  It does 
not treat the contribution of the three-nucleon
force to the quasiparticle interaction but sets a reliable framework for 
future calculations employing also the leading-order chiral three-nucleon 
interaction \cite{holt11}. In the present work, we therefore seek to identify deficiencies 
that remain when only two-nucleon forces are included in the calculation of the 
quasiparticle interaction.

The paper is organized as follows. In Section \ref{qpisec} we describe the microscopic
approach to Landau's Fermi liquid theory and relate the $L=0$ and $L=1$ Landau
parameters to various nuclear matter observables. We then describe in detail our 
complete calculation of the quasiparticle interaction to second order in perturbation
theory. In Section \ref{calres} we first apply our scheme to analytically-solvable
model interactions (scalar-isoscalar boson exchange and modified pion exchange) 
in order to assess the numerical accuracy. We then 
employ realistic low-momentum nucleon-nucleon interactions and make contact
to experimental quantities through the Landau parameters. The paper ends with a 
summary and outlook.

\section{Nuclear quasiparticle interaction}
\label{qpisec}
\subsection{Landau parameters and nuclear observables}

The physics of `normal' Fermi liquids at low temperatures is governed by the
properties and interactions of quasiparticles, as emphasized by Landau in the 
early 1960's. Since quasiparticles are 
well-defined only near the Fermi surface ($|\vec p\,| = k_F$) where they are long-lived, Landau's 
theory is valid only for low-energy excitations about the interacting
ground state. The quantity of primary importance in the theory is the 
interaction energy between two quasiparticles, which can be obtained by
functionally differentiating the ground-state energy density twice with respect 
to the quasiparticle densities:
\begin{equation}
{\cal F}({\vec p}_1 s_1 t_1; {\vec p}_2 s_2 t_2)
= \left . \frac{\delta^2 {\cal E}}{\delta \tilde n_1 \delta \tilde n_2} \right
 |_{\tilde n_1= \tilde n_2=0} \, ,
\label{qpi}
\end{equation}
where $s_{1,2}=\pm 1/2$ and $t_{1,2}=\pm 1/2$ are spin and isospin quantum numbers.
The general form of the central part of the quasiparticle interaction in nuclear matter
excluding tensor components, etc., is given by
\begin{equation}
{\cal F}({\vec p}_1, {\vec p}_2) =
f({\vec p}_1,{\vec p}_2) + f^\prime({\vec p}_1,{\vec p}_2) {\vec \tau}_1 \cdot
{\vec \tau}_2 + \left [g({\vec p}_1,{\vec p}_2) + g^\prime({\vec p}_1,{\vec p}_2) 
{\vec \tau}_1 \cdot {\vec \tau}_2\right ] {\vec \sigma}_1 \cdot {\vec \sigma}_2 \, ,
\label{ffunction}
\end{equation}
where $\vec \sigma_{1,2}$ and $\vec \tau_{1,2}$ are respectively the spin and isospin operators 
of the two nucleons on the Fermi sphere $|\vec p_1\,| = |\vec p_2\,| = k_F$. 
For notational simplicity we have dropped the dependence on the quantum numbers 
$s_{1,2}$ and $t_{1,2}$, which is introduced through the matrix elements of the 
operators: $\vec \sigma_1 \cdot \vec \sigma_2 \to 4 s_1s_2 = \pm 1$ and
$\vec \tau_1 \cdot \vec \tau_2 \to 4 t_1t_2 = \pm 1$. As it stands in eq.\ 
(\ref{ffunction}), the quasiparticle interaction is defined for any nuclear 
density $\rho=2k_F^3/3\pi^2$, but the quantities of physical interest result
at nuclear matter saturation density $\rho_0 \simeq 0.16$\,fm$^{-3}$ 
(corresponding to $k_F = 1.33$\,fm$^{-1}$). 
For two quasiparticles on the Fermi surface $|\vec p_1\,| = |\vec p_2\,| = k_F$, the 
remaining angular 
dependence of their interaction can be expanded in Legendre polynomials of
$\cos \theta =\hat p_1 \cdot \hat p_2$:
\begin{equation}
X({\vec p}_1,{\vec p}_2) = \sum_{L=0}^\infty X_L P_L(\mbox{cos } \theta), \hspace{.1in}
\end{equation}
where $X$ represents $f, f^\prime, g,$ or $g^\prime$, and the angle $\theta$ is 
related to the relative momentum $p = \frac{1}{2} |{\vec p}_1 - {\vec p}_2|$ through the relation
\begin{equation}
p = k_F\, {\rm sin}\, \frac{\theta}{2}\, .
\end{equation}
It is conventional to factor out from the quasiparticle interaction 
the density of states per unit energy and volume at the Fermi surface, $N_0=2M^*k_F/\pi^2$, 
where $M^*$ is the nucleon effective mass (see eq.\ (\ref{effmasseq})) and $k_F =1.33$\,fm$^{-1}$. 
This enables one to introduce an equivalent set of
dimensionless Fermi liquid parameters $F_L, G_L, F_L^\prime,$ and $G_L^\prime$ 
through the relation
\begin{equation}
{\cal F}({\vec p}_1, {\vec p}_2)=\frac{1}{N_0}\sum_{L=0}^\infty \left [
F_L + F^\prime_L \vec \tau_1 \cdot
\vec \tau_2 + (G_L + G^\prime_L \vec \tau_1 \cdot \vec \tau_2) \vec \sigma_1 \cdot \vec \sigma_2 \right ]
P_L({\rm cos}\, \theta).
\label{ffunction2}
\end{equation}
Provided the above series converges quickly in $L$, the interaction between two
quasiparticles on the Fermi surface is governed by just a few constants which 
can be directly related to a number of observable quantities as we now discuss.

The quasiparticle effective mass $M^*$ is related to the 
slope of the single-particle potential at the Fermi surface and can be obtained from 
the Landau parameter $f_1$ by invoking Galilean invariance. The relation is found
to be
\begin{equation}
\frac{1}{M^*} = \frac{1}{M_N}-\frac{2k_F}{3 \pi^2} f_1 ,
\label{effmasseq}
\end{equation}
where $M_N = 939$ MeV is the free nucleon mass.
The compression modulus ${\cal K}$ of symmetric nuclear matter can be obtained from
the isotropic ($L=0$) spin- and isospin-independent component of the 
quasiparticle interaction
\begin{equation}
{\cal K}=\frac{3k_F^2}{M^*}\left (1+F_0\right ).
\end{equation}
The compression modulus of infinite nuclear matter cannot be measured directly, 
but its value ${\cal K}=250 \pm 50$ MeV can be estimated from theoretical 
predictions of giant monopole resonance energies 
in heavy nuclei \cite{blaizot,youngblood,ring}. The nuclear symmetry energy $\beta$
can be computed from the isotropic spin-independent part of the isovector interaction:
\begin{equation}
\beta = \frac{k_F^2}{6M^*}(1+F_0^\prime).
\end{equation}
Global fits of nuclear masses with semi-empirical binding energy formulas provide 
an average value for the symmetry energy of $\beta = 33 \pm 3$ MeV over densities in the
vicinity of saturated nuclear matter \cite{danielewicz,steiner}. 
The quasiparticle interaction provides also a link to the properties of 
single-particle and collective excitations. In particular, the orbital  
$g$-factor for valence nucleons (i.e., quasiparticles on the Fermi surface) is 
different by the amount $\delta g_l$ from that of a free nucleon:
\cite{migdal2}: 
\begin{equation}
g_l = \frac{1+\tau_3}{2} + \frac{F^\prime_1-F_1}{6(1+F_1/3)}\tau_3 \equiv
\frac{1+\tau_3}{2} + \delta g_l \tau_3.
\label{aog}
\end{equation}
One possible mechanism for the anomalous orbital $g$-factor $\delta g_l$ are 
meson exchange currents \cite{miyazawa,brown80}, 
which arise in the isospin-dependent components of the nucleon-nucleon interaction.
According to eq.\ (\ref{aog}), the renormalized isoscalar and isovector orbital $g$-factors
are 
\begin{eqnarray}
g_l^{(is)}&=&\frac{1}{2}\left (g^{(p)}_l+g^{(n)}_l \right )=\frac{1}{2} \, , \nonumber \\
g_l^{(iv)}&=&\frac{1}{2}\left (g^{(p)}_l-g^{(n)}_l \right )= \frac{M_N}{2M^*}\left ( 1 + \frac{F^\prime_1}{3} \right ) =\frac{1}{2}+\delta g_l
\end{eqnarray}
are different. The former receives no correction, while the 
latter is sizably enhanced by the (reduced) effective mass $M^*$ as well as by the 
(positive) Landau parameter $F^\prime_1$. It receives a large contribution from 
one-pion exchange. 

Nuclear matter 
allows for a rich variety of collective states, including density 
(breathing mode), spin (magnetic dipole mode), isospin (giant dipole mode), and 
spin-isospin (giant Gamow-Teller mode) excitations. As previously discussed, the 
breathing mode is governed by the incompressibility ${\cal K}$ of nuclear matter \cite{youngblood}. 
The energy of the (isovector) giant dipole mode is correlated with the nuclear symmetry energy
$\beta$ \cite{trippa}, while the dipole sum rule \cite{brown80} 
\begin{equation}
\int_0^{\omega_{\rm max}} d\omega \, \sigma(E1) = \frac{2\pi \alpha}{M_N} \frac{NZ}{A} (1+2\delta g_l)\,
\end{equation}
is connected to the anomalous orbital $g$-factor $\delta g_l$ with $\alpha = 1/137$.
Experimental results \cite{nolte} 
are consistent with a value of the anomalous orbital $g$-factor of
$\delta g_l\simeq 0.23 \pm 0.03$. Finally, the
giant Gamow-Teller resonance has been widely studied due to its connection to the
nuclear spin-isospin response function and for ruling out pion condensation in
moderately-dense nuclear matter. An analysis of the 
experimental excitation energies and transition strengths \cite{gaarde83,ericson88,suzuki99} leads to a 
value for the parameter
\begin{equation}
g_{NN}^\prime \simeq 0.6-0.7 ,
\end{equation}
which is used to model the spin-isospin interaction in nuclei as a zero-range contact
interaction.  As a convention it is related to the dimensionless Landau parameter $G_0^\prime$ by
\begin{equation}
G^\prime_0 = N_0 \frac{g_{\pi N}^2}{4M_N^2} g_{NN}^\prime \, ,
\end{equation}
where $g_{\pi N} \simeq 13.2$ is the strong $\pi N$ coupling constant.
It is well-known that the giant Gamow-Teller resonances receive
important contributions from the coupling to $\Delta$-hole excitations \cite{suzuki99}.
Such dynamical effects due to non-nucleonic degrees of freedom are reflected in the 
leading-order, $2\pi$-exchange three-nucleon interaction
to be included in future work \cite{holt11}.

\subsection{Quasiparticle interaction at second order}
\label{qp2n}

\begin{figure}
\begin{center}
\includegraphics[height=4cm]{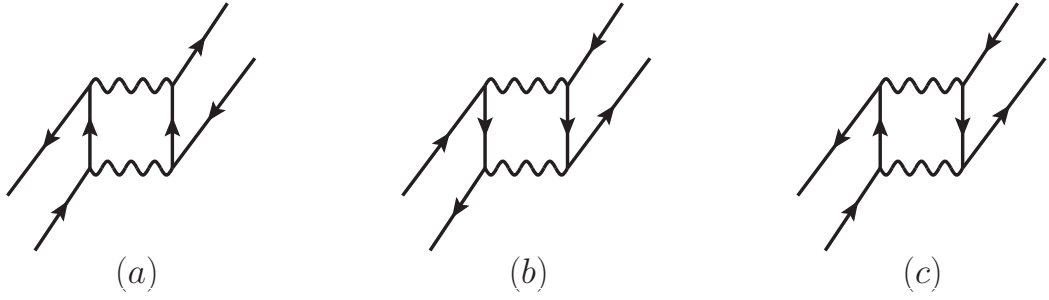}
\end{center}
\vspace{-.5cm}
\caption{Diagrams contributing to the second-order quasiparticle interaction (exchange
terms omitted): (a)
particle-particle diagram, (b) hole-hole diagram, and (c) particle-hole diagram.}
\label{pphhph}
\end{figure}

Expanding the energy density to second-order in the (Goldstone) linked-diagram
expansion and differentiating twice with respect to the nucleon distribution
function, one obtains for the first two contributions to the 
quasiparticle interaction
\begin{equation}
{\cal F}^{(1)}({\vec p}_1 s_1 t_1; {\vec p}_2 s_2 t_2)
 = \langle {\vec p}_1 s_1 t_1; {\vec p}_2 s_2 t_2 | 
\bar V | {\vec p}_1 s_1 t_1; {\vec p}_2 s_2 t_2 \rangle
\equiv \langle 12 | \bar V | 12 \rangle
\label{order1qp}
\end{equation}
and
\begin{eqnarray}
{\cal F}^{(2)} ({\vec p}_1 s_1 t_1; {\vec p}_2 s_2 t_2)
&=& \frac{1}{2} \sum_{34} \frac{|\langle 12 | \bar V | 34 \rangle|^2 (1-n_3)(1-n_4)}
{\epsilon_1 + \epsilon_2 - \epsilon_3 - \epsilon_4} \nonumber \\
 &+& \frac{1}{2} \sum_{34} \frac{|\langle 12 | \bar V | 34 \rangle |^2 n_3 n_4 }
{\epsilon_3 + \epsilon_4 - \epsilon_1 - \epsilon_2} 
 - 2 \sum_{34} \frac{|\langle 13 | \bar V | 24 \rangle |^2 n_3 (1-n_4)}
{\epsilon_1 + \epsilon_3 - \epsilon_2 - \epsilon_4}.
\label{order2qp}
\end{eqnarray}
In eqs.\ (\ref{order1qp}) and (\ref{order2qp}) the quantity $\bar V$ 
denotes the antisymmetrized two-body potential (with units of fm$^2$) given by 
$\langle p l S J T | \bar V | p^\prime l^\prime S J T \rangle = (1-(-1)^{L+S+T}) 
\langle p l S J T | V | p^\prime l^\prime S J T \rangle$ in the partial wave 
basis, and in eq.\ (\ref{order2qp}) the summation is over 
intermediate-state momenta, spins and isospins. We specify our sign and normalization 
conventions through the perturbative relation between 
diagonal two-body matrix elements and phase shifts: $\tan \delta_{lSJ}(p)
= -p M_N \langle plSJT | V | plSJT \rangle /4\pi$.
The first-order term of eq.\ (\ref{order1qp}) is just the diagonal matrix element of the 
antisymmetrized two-body interaction, while 
the second-order term (eq.\ (\ref{order2qp})) has been separated into particle-particle, hole-hole, 
and particle-hole terms depicted diagrammatically in Fig.\ \ref{pphhph}. The 
distribution function $n$ is the usual step function for the nuclear matter ground state:
\begin{equation}
n_{\vec k} = \left \{ \begin{array}{cl} 
1 & {\rm for} \hspace{.1in} | \vec k\,| \leq k_F \\
0 & {\rm for} \hspace{.1in} |\vec k \,| > k_F  \end{array} \right . .
\end{equation}

In the following, we discuss the general evaluation of eqs.\ 
(\ref{order1qp}) and (\ref{order2qp}) for interactions given in the
partial-wave basis. We first define the spin-averaged quasiparticle interaction
$\bar {\cal F}(pST)$, which is obtained from the full quasiparticle 
interaction by averaging over the spin-substates:
\begin{equation}
\bar {\cal F}(pST) = \frac{1}{2S+1}\sum_{m_s} {\cal F}(pSm_sTT_z),
\end{equation}
where $p = \frac{1}{2} | \vec p_1 - \vec p_2 \,| = k_F \sin \theta/2$, and in
${\cal F}(pSm_sTT_z)$
the spins and isospins of the two quasiparticles are coupled to total
spin $S=0,1$ and total isospin $T=0,1$. We take an isospin-symmetric two-body potential 
and thus the quasiparticle interaction is independent of $T_z$.
The first-order contribution to the central part of the quasiparticle interaction is 
then obtained by summing over the allowed partial wave matrix elements:
\begin{equation}
\bar {\cal F}^{(1)}(pST) = \frac{1}{2S+1} \sum_{J,l} (2J+1) \left \langle 
p lSJT \right | \bar V \left | p lSJT \right \rangle \, .
\label{qpdec1}
\end{equation}
Note that
there is an additional factor of $4\pi$ in eq.\ (41) in ref.\ \cite{schwenk02} and 
eq.\ (28) in ref.\ \cite{holt07} due to a different normalization convention.
From eq.\ (\ref{qpdec1}) we can project out the individual components
of the quasiparticle interaction using the appropriate linear combinations of 
$\bar{\cal F}(pST)$ with $S=0,1$ and $T=0,1$:
\begin{eqnarray}
f(p) &=&\; \; \, \frac{1}{16}\bar {\cal F}^{(1)}(p00) + \frac{3}{16}\bar {\cal F}^{(1)}(p01) + 
\frac{3}{16}\bar {\cal F}^{(1)}(p10) + \frac{9}{16}\bar {\cal F}^{(1)}(p11) \nonumber \\
g(p) &=& -\frac{1}{16}\bar {\cal F}^{(1)}(p00) - \frac{3}{16}\bar {\cal F}^{(1)}(p01) + 
\frac{1}{16}\bar {\cal F}^{(1)}(p10) + \frac{3}{16}\bar {\cal F}^{(1)}(p11) \nonumber \\
f^\prime (p) &=& -\frac{1}{16}\bar {\cal F}^{(1)}(p00) + \frac{1}{16}\bar {\cal F}^{(1)}(p01) 
- \frac{3}{16}\bar {\cal F}^{(1)}(p10) + \frac{3}{16}\bar {\cal F}^{(1)}(p11) \nonumber \\
g^\prime (p) &=& \; \; \, \frac{1}{16}\bar {\cal F}^{(1)}(p00) - \frac{1}{16}\bar {\cal F}^{(1)}(p01) -
\frac{1}{16}\bar {\cal F}^{(1)}(p10) + \frac{1}{16}\bar {\cal F}^{(1)}(p11).
\label{lincomb}
\end{eqnarray}
The leading-order expressions, eqs.\ (\ref{qpdec1}) and (\ref{lincomb}), give the full 
$p$-dependence (i.e., angular dependence) of the quasiparticle interaction, 
and therefore one can project out the 
density-dependent Landau parameters for arbitrary $L$:
\begin{equation}
X_L = 2(2L+1) \int_0^{k_F} dp \frac{p}{k_F^2} X(p) P_L(1-2p^2/k_F^2).
\end{equation}

For the second-order contributions to the quasiparticle interaction, the complete 
$p$-dependence is in general not easily obtained (e.g., for the particle-hole term). 
We instead compute the Landau parameters for each $L$ separately, choosing the 
total momentum vector to be aligned with the $z$-axis. In the following, the two 
quasiparticle momenta are labeled $\vec p_1$ and
$\vec p_2$, while the intermediate-state momenta are labeled $\vec k_3$ and
$\vec k_4$. For the particle-particle contribution one finds
\begin{eqnarray}
&&{\cal F}^{(2)pp}_L(Sm_sT) = \frac{2L+1}{4 \pi^2 k_F^2} \sum_{\substack {l_1l_2l_3l_4mm^\prime \\ 
m_s^\prime JJ^\prime M}} \int_0^{k_F} dp \, p \int_p^{\infty} dq \, q^2 N(l_1ml_2m^\prime l_3m l_4m^\prime) 
P_{l_1}^m(0) P_{l_3}^m(0) \nonumber \\
&& \hspace{.3in} \times   \frac{M_N}{p^2-q^2} i^{l_2+l_3-l_1-l_4} {\cal C}^{JM}_{l_1msm_s} 
{\cal C}^{JM}_{l_2m^\prime sm_s^\prime} 
{\cal C}^{J^\prime M}_{l_3msm_s} {\cal C}^{J^\prime M}_{l_4m^\prime sm_s^\prime}  
\int^{{\rm min}\{ x_0,1 \}}_{{\rm max}\{ -x_0,-1 \}} 
d \cos  \theta_q \, P_{l_2}^{m^\prime}(\cos \theta_q) P_{l_4}^{m^\prime}(\cos \theta_q) \nonumber \\
&& \hspace{.3in} \times  \langle pl_1SJMT| \bar V | ql_2SJMT \rangle
\langle ql_4SJ^\prime MT | \bar V | pl_3SJ^\prime MT \rangle P_L(1-2p^2/k_F^2),
\label{pp2nd}
\end{eqnarray}
where $P_l^m$ are associated Legendre functions, $\vec p = (\vec p_1 -\vec p_2)/2$, 
$\vec q = (\vec k_3 - \vec k_4)/2$, 
$x_0 = (q^2-p^2)/2q\sqrt{k_F^2-p^2}$ and $N(l_1ml_2m^\prime l_3m l_4m^\prime) = N_{l_1}^m
N_{l_2}^{m^\prime} N_{l_3}^m N_{l_4}^{m^\prime}$ with $N_l^m = \sqrt{(2l+1)(l-m)!/(l+m)!}$.
Similarly, for the hole-hole diagram one obtains
\begin{eqnarray}
&& {\cal F}^{(2)hh}_L(Sm_sT) = \frac{2L+1}{4 \pi^2 k_F^2} \sum_{\substack {l_1l_2l_3l_4mm^\prime 
\\ m_s^\prime JJ^\prime M}}
\int_0^{k_F} dp \, p \int_0^p dq \, q^2 N(l_1ml_2m^\prime l_3m l_4m^\prime) P_{l_1}^m(0)
P_{l_3}^m(0) \nonumber \\
&& \hspace{.3in} \times \frac{M_N}{q^2-p^2} i^{l_2+l_3-l_1-l_4} {\cal C}^{JM}_{l_1msm_s} 
{\cal C}^{JM}_{l_2m^\prime sm_s^\prime} 
{\cal C}^{J^\prime M}_{l_3msm_s} {\cal C}^{J^\prime M}_{l_4m^\prime sm_s^\prime}  
\int^{{\rm min}\{ -x_0,1 \}}_{{\rm max}\{ x_0,-1 \}} 
d\cos \theta_q \, P_{l_2}^{m^\prime}(\cos \theta_q) P_{l_4}^{m^\prime}(\cos \theta_q)  \nonumber \\
&& \hspace{.3in} \times \langle pl_1SJMT | \bar V | ql_2SJMT \rangle
\langle ql_4SJ^\prime MT | \bar V | pl_3SJ^\prime MT \rangle P_L(1-2p^2/k_F^2).
\label{hh2nd}
\end{eqnarray}
Averaging over the spin substates and employing eq.\ (\ref{lincomb}) with the substitution 
$\bar {\cal F}^{(1)} \rightarrow \bar {\cal F}^{(2)}$ again yields the individual spin and isospin
components of the quasiparticle interaction. The evaluation of the particle-hole 
diagram proceeds similarly; however, in this case the coupling of the two quasiparticles to total spin
(and isospin) requires an additional step. Coupling to states with $m_s=0$ is achieved by
taking the combinations (neglecting isospin for simplicity)
\begin{eqnarray}
&&{\cal F}^{(2)ph}({\vec p}_1 {\vec p}_2; S=1/2 \pm 1/2, m_s=0) =  - 2 \sum_{34} \left [ 
\langle \vec p_1 \vec k_3 \uparrow s_3 | \bar V | \vec p_2 \vec k_4 \downarrow s_4 \rangle 
\langle \vec p_2 \vec k_4 \downarrow s_4 | \bar V | \vec p_1 \vec k_3 \uparrow s_3 \rangle \right . 
\nonumber \\
&&\pm \left . \langle \vec p_1 \vec k_3 \uparrow s_3 | \bar V | \vec p_2 \vec k_4 \uparrow s_4 \rangle 
\langle \vec p_2 \vec k_4 \downarrow s_4 | \bar V | \vec p_1 \vec k_3 \downarrow s_3 \rangle \right ]
\frac{n_3 (1-n_4)}{\epsilon_1 + \epsilon_3 - \epsilon_2 - \epsilon_4}.
\label{phcomb}
\end{eqnarray}
We provide the expression for ${\cal F}^{(2)ph}_0$ (corresponding to $L=0$) 
in uncoupled quasiparticle spin and isospin states, 
appopriate for evaluating the first term in eq.\ (\ref{phcomb}), which can be easily generalized in order 
to obtain the second term. We find
\begin{eqnarray}
&& {\cal F}^{(2)ph}_0(s_1 s_2 t_1 t_2) = -\frac{1}{\pi^2 k_F^2} 
\sum_{\substack {s_3 s_4 t_3 t_4\\ l_1l_2l_3l_4mm^\prime \\ 
SJMTS^\prime J^\prime T^\prime}}
\int_0^{2k_F} dp^\prime \int_{k_F-p^\prime/2}^{\sqrt{k_F^2-{p^\prime}^2/4}}  dp \, p 
\int_{\sqrt{k_F^2-{p^\prime}^2/4}}^{k_F+p^\prime/2} dq \, q  \,N(l_1ml_2m^\prime l_3m l_4m^\prime)   \nonumber \\
&& \times \frac{M_N}{p^2-q^2}  P_{l_1}^m(\cos \theta_p) P_{l_3}^m(\cos \theta_p) 
P_{l_2}^{m^\prime} (\cos \theta_q) P_{l_4}^{m^\prime}(\cos \theta_q) i^{l_2+l_3-l_1-l_4} 
{\cal C}^{JM}_{l_1mSm_s} {\cal C}^{JM}_{l_2m^\prime Sm_s^\prime} 
{\cal C}^{J^\prime M}_{l_3mS^\prime m_s} {\cal C}^{J^\prime M}_{l_4m^\prime S^\prime m_s^\prime} \nonumber \\
&& \times {\cal C}(s_1s_2s_3s_4){\cal C}(t_1t_2t_3t_4)\langle pl_1SJMT | \bar V | ql_2SJMT \rangle
\langle ql_4S^\prime J^\prime MT^\prime | \bar V | pl_3S^\prime J^\prime MT^\prime \rangle ,
\label{ph2nd}
\end{eqnarray}
where now $\vec p = (\vec p_1 -\vec k_3)/2$, $\vec q = (\vec p_2 -\vec k_4)/2$, 
and the total momentum $\vec p^{\, \prime}= \vec p_1 + \vec k_3 = \vec p_2 + \vec k_4$. 
The angle $\theta_p$ between $\vec p$ and $\vec p^{\, \prime}$ is fixed (via $|\vec p_1\, | = k_F$ 
together with $\vec p_1 = \vec p^{\, \prime}/2 + \vec p$\,)
by the relation $pp^\prime \cos \theta_p = k_F^2-p^2-{p^\prime}^2/4$, and analogously for the angle
$\theta_q$ between $\vec q$ and $\vec p^{\, \prime}$.
The combination of spin Clebsch-Gordan coefficients that arises in the above expression 
is denoted by
\begin{equation}
{\cal C}(s_1s_2s_3s_4) = {\cal C}^{Sm_s}_{\frac{1}{2}s_1 \frac{1}{2}s_3} 
{\cal C}^{Sm_s^\prime}_{\frac{1}{2}s_2 \frac{1}{2}s_4} {\cal C}^{S^\prime m_s}_{\frac{1}{2}s_1 \frac{1}{2}s_3}
{\cal C}^{S^\prime m_s^\prime}_{\frac{1}{2}s_2 \frac{1}{2}s_4},
\end{equation}
and likewise for the combination of isospin Clebsch-Gordan coefficients. In computing 
the particle-hole term for $L>0$, we use
\begin{equation}
P_L(\hat p_1 \cdot \hat p_2) = P_L \left (\frac{1}{4k_F^2} {p^{\, \prime}}^2 + \frac{1}{2k_F^2}
p^{\, \prime}(p\cos \theta_p + q\cos \theta_q) + \frac{p q}{k_F^2} \cos \theta_{pq} \right ),
\end{equation}
and employ the addition theorem for spherical harmonics to write $\cos \theta_{pq}$ in terms of 
$\cos \theta_p$, $\cos \theta_q$ and an azimuthal angle $\phi$. The involved integral $\int_0^{2\pi} d\phi$ 
gives different selection rules for the $m,m^\prime$ values of the associated Legendre functions 
in eq.\ (\ref{ph2nd}).
In deriving eqs.\ (\ref{pp2nd})--(\ref{ph2nd}), 
we have assumed that the intermediate-state energies in eq.\ (\ref{order2qp}) are those of free particles:
$\epsilon_k = \vec k^2/2M_N$. Later we will include the first-order correction to the dispersion relation
arising from the in-medium self-energy, which leads to the substitution $M_N \rightarrow M^*$ in the
above equations.

\section{Calculations and results}
\label{calres}

\subsection{Prelude: One boson exchange interactions as test cases}

The numerical computation of the quasiparticle interaction at second order is
obviously quite intricate, and truncations in the number of included 
partial waves and in the momentum-space integrations are necessary. In such
a situation it is very helpful to have available analytical results 
for simple model interactions in order to test the accuracy of the numerical
calculations. For that purpose we derived in this subsection analytical expressions 
for the quasiparticle interaction up to second order arising from (i) massive 
scalar-isoscalar boson exchange and (ii) pion exchange modified by squaring 
the static propagator. We omit all technical details of these calculations which can 
be found (for $L=0$) in ref.\ \cite{kaiser06} for the case of tree-level and one-loop 
(i.e., second-order) pion-exchange. In the present treatment the second-order quasiparticle interaction 
is organized differently than in eq.\ (\ref{order2qp}). The explicit decomposition 
of the in-medium nucleon propagator into a particle and hole propagator is
replaced by the sum of a ``vacuum'' and ``medium insertion'' component:
\begin{eqnarray}
G(p_0,\vec p\,)&=&i\left (\frac{\theta(|\vec p\,|-k_F)}{p_0-\vec p^{\,\,2}/2M_N+i\epsilon}
+\frac{\theta(k_F - |\vec p\,|)}{p_0-\vec p^{\,\,2}/2M_N-i\epsilon}\right ) \nonumber \\
&=& \frac{i}{p_0-\vec p^{\,\,2}/2M_N+i\epsilon}-2\pi \delta(p_0-\vec p^{\,\,2}/2M_N)
\theta(k_F - |\vec p\,|),
\end{eqnarray}
and the organization is now in the number of medium insertions rather than
in terms of particle and hole intermediate states. The central parts of the quasiparticle
interaction are constructed for any $L$ through an angle-averaging procedure 
\begin{eqnarray}
{\cal F}_L(k_F) &=& \frac{2L+1}{(4\pi)^2} \int d\Omega_1 d\Omega_2 \langle \vec p_1
\vec p_2 | V_{\rm eff} | \vec p_1 \vec p_2 \rangle P_L(\hat p_1 \cdot \hat p_2) \nonumber \\
&=& f_L(k_F) + g_L(k_F) \vec \sigma_1 \cdot \vec \sigma_2 + f^\prime_L(k_F)
\vec \tau_1 \cdot \vec \tau_2 + g^\prime _L(k_F) \vec \sigma_1 \cdot \vec \sigma_2\,
\vec \tau_1 \cdot \vec \tau_2\, .
\end{eqnarray}
In this equation $V_{\rm eff}$ represents the effective model interaction computed up 
to one-loop order (second order).

\begin{figure}
\begin{center}
\includegraphics[height=3cm]{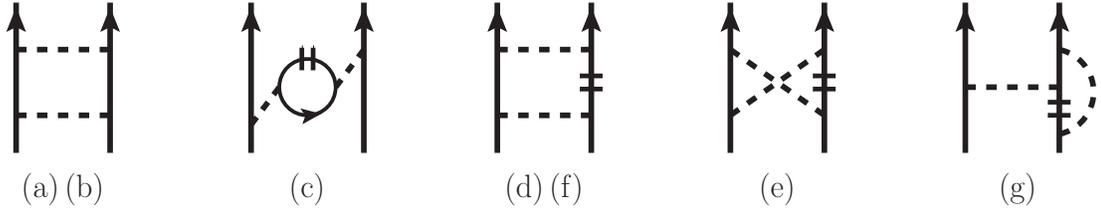}
\end{center}
\vspace{-.5cm}
\caption{Diagrammatic contributions to the second-order quasiparticle interaction in nuclear
matter organized in the number of medium insertions (symbolized by the short double lines).
Crossed diagrams and additional reflected diagrams are not shown. The labels (b) and (f) refer
to the crossed terms of (a) and (d).}
\label{qp2norbert}
\end{figure}

We first consider as a generic example the exchange of a scalar-isoscalar boson with mass $m_s$ and 
coupling constant $g_s$ (to the nucleon). In momentum and coordinate space it
gives rise to central potentials of the form
\begin{equation} 
V_C(q)= -\frac{g_s^2}{m_s^2+q^2} \,,\qquad \widetilde V_C(r) = 
-\frac{g_s^2}{4\pi}\frac{e^{-m_s r}}{r}\, .
\label{scisop}
\end{equation}
For the 
first-order contributions to the $L=0,1$ Landau parameters one finds
\begin{equation}  {\cal F}^{(1)}_0(k_F) ={g_s^2\over m_s^2}  \bigg[-1+(1+
\boldsymbol\sigma)(1+\boldsymbol\tau)\, {\ln(1+4u^2)\over 16u^2}\bigg] \,,
\label{sciso1}
\end{equation}
\begin{equation}  {\cal F}^{(1)}_1(k_F) 
=(1+\boldsymbol\sigma)(1+\boldsymbol\tau
){3g_s^2\over 32m_s^2u^4} \Big[(1+2u^2) \ln(1+4u^2)-4u^2\Big] \,, 
\end{equation}
where $\boldsymbol\sigma= \vec \sigma_1 \cdot\vec
\sigma_2$ and $\boldsymbol\tau= \vec \tau_1 \cdot
\vec\tau_2$ are short-hand notations for the spin-spin and isospin-isospin operators.
The dimensionless variable $u=k_F/m_s$ denotes the ratio of the Fermi momentum $k_F$ to 
the scalar boson mass $m_s$. Note that in this approach both direct and crossed diagrams 
can contribute. The crossed diagrams have to be multiplied by the negative product of the spin- and 
isospin-exchange operators $-(1+\boldsymbol \sigma)(1+\boldsymbol\tau)/4$.
At second order there are five classes of diagrammatic contributions, shown in Fig.\ 
\ref{qp2norbert},
to the quasiparticle interaction. The direct terms from iterated (second order) 
boson exchange, see Fig.\ \ref{qp2norbert}(a), read
\begin{equation}  {\cal F}^{(2a)}_0(k_F) = -{g_s^4 M_N \over 32 \pi m_s^3} \,
{\ln(1+4u^2)\over u^2} \,, \end{equation}
\begin{equation}  {\cal F}^{(2a)}_1(k_F) = {3g_s^4 M_N \over 64 \pi m_s^3 
u^4}
\Big[4u^2-(1+2u^2)\ln(1+4u^2)\Big] \,, \end{equation}
whereas the corresponding crossed terms (b) have the form
\begin{equation}  {\cal F}^{(2b)}_0(k_F) = (1+\boldsymbol\sigma)(1+
\boldsymbol\tau){g_s^4 M_N \over 16 \pi m_s^3}\int_0^u\!\!dx \, {\arctan 2x-
\arctan x\over u^2(1+2x^2)} \,, \end{equation}
\begin{equation}  {\cal F}^{(2b)}_1(k_F) = (1+\boldsymbol\sigma)(1+
\boldsymbol\tau){3g_s^4 M_N \over 16\pi m_s^3} \int_0^u\!\!dx \, {u^2-2x^2
\over u^4 (1+2x^2)}\big[\arctan 2x-\arctan x\big] \, . \end{equation}
The coupling of the exchanged boson to nucleon-hole states, Fig.\ 
\ref{qp2norbert}(c), gives rise to nonvanishing crossed terms which read
\begin{equation} {\cal F}^{(2c)}_0(k_F) = (1+\boldsymbol\sigma)(1+
\boldsymbol\tau){g_s^4 M_N \over 4\pi^2 m_s^3 u^2}\int_0^u \!\!dx\,{1\over
(1+4x^2)^2} \bigg[ 2u x +(u^2-x^2) \ln{u+x \over u-x} \bigg] \,, 
\end{equation}
\begin{equation} {\cal F}^{(2c)}_1(k_F) = (1+\boldsymbol\sigma)(1+
\boldsymbol\tau){3g_s^4 M_N \over 4\pi^2 m_s^3 u^4}\int_0^u 
\!\!dx\,{u^2-2x^2\over
(1+4x^2)^2} \bigg[ 2u x +(u^2-x^2) \ln{u+x\over u-x} \bigg] \, .
\end{equation}
Pauli blocking occurs in the planar- and crossed-box diagrams, Fig.\ 
\ref{qp2norbert}(d)--(e), and for the sum of their direct terms one finds the forms
\begin{equation} {\cal F}^{(2d+2e)}_0(k_F) ={2g_s^4 M_N \over \pi^2 m_s^3 u^2}
\int_0^u\!\!dx \,{x\over (1+4x^2)^2} \bigg[ u\ln{u \over u-x} +x
\ln{u-x \over x}    \bigg] \,, \end{equation}
\begin{eqnarray} {\cal F}^{(2d+2e)}_1(k_F) &=&{g_s^4 M_N \over \pi^2 m_s^3 
u^4}
\int_0^u\!\!dx \,{x\over (1+4x^2)^2} \bigg[ 2ux(x-u)+2u^3\ln{u \over u-x}
\nonumber \\ && \qquad \qquad \quad +3x(u^2-x^2)\ln{u^2-x^2 \over x^2}+2x^3
\ln{u-x \over x} \bigg] \, . \end{eqnarray}
On the other hand, the crossed terms of the planar-box diagram with Pauli
blocking, see Fig.\ \ref{qp2norbert}(f) yield
\begin{eqnarray} {\cal F}^{(2f)}_0(k_F) &=&-(1+\boldsymbol\sigma)(1+
\boldsymbol\tau){g_s^4M_N \over 4\pi^2 m_s^3
u^2} \int_0^u \!\!dx \,{x \over 1+4x^2}\nonumber \\ && \times 
\int_0^{u-x}\!\!\!
dy\,{1\over \sqrt{R}} \ln{u \sqrt{R} +(1-4x y)(x-y) \over u \sqrt{R} 
+(4x y-1)
(x-y)} \,, \end{eqnarray}
\begin{eqnarray} {\cal F}^{(2f)}_1(k_F) &=&-(1+\boldsymbol\sigma)(1+
\boldsymbol\tau){3g_s^4M_N \over 64\pi^2 m_s^3
u^4} \int_0^u \!\!dx \Bigg\{\ln^2{1+(u+x)^2\over 1+(u-x)^2} +{8x \over 
1+4x^2}
\bigg[ u \ln{4(u-x)\over u+x}\nonumber \\ && -x \ln{u^2-x^2\over x^2}+
\int_0^{u-x}\!\!\! dy\,{1+2u^2-4x^2\over \sqrt{R}} \ln{u \sqrt{R} +(1-4x 
y)(x-y)
\over u \sqrt{R} +(4x y-1)(x-y)}\bigg] \Bigg\}\,, \end{eqnarray}
with the auxiliary polynomial $R=4u^2+(4x^2-1)(4y^2-1)$. Finally, the density-dependent 
vertex correction to one-boson exchange, Fig.\ \ref{qp2norbert}(g), 
provides a nonzero contribution only in the crossed diagram. The
corresponding expressions for the $L=0,1$ Landau parameters read
\begin{equation} {\cal F}^{(2g)}_0(k_F) = -(1+\boldsymbol\sigma)(1+
\boldsymbol\tau){g_s^4 M_N \over 16\pi^2 m_s^3
u^2} \int_0^u \!\! dx \, {\ln(1+4x^2) \over \sqrt{1+4u^2-4x^2}} \ln { ( 
u\sqrt{
1+4u^2-4x^2} +x )^2 \over(1+4u^2)(u^2-x^2)} \, , \end{equation}

\begin{eqnarray} {\cal F}^{(2g)}_1(k_F) &=& (1+\boldsymbol\sigma)(1+
\boldsymbol\tau){3g_s^4 M_N \over 32\pi^2
m_s^3 u^4} \int_0^u \!\! dx  \ln(1+4x^2) \bigg\{\ln{u+x\over u-x} 
\nonumber \\
&& -{1+2u^2 \over \sqrt{1+4u^2-4x^2}} \ln { ( u\sqrt{1+4u^2-4x^2} +x )^2 
\over
(1+4u^2)(u^2-x^2)} \bigg\}\, . 
\label{sc2g}
\end{eqnarray}
Since at most double integrals over well-behaved functions are involved 
in the expressions in eqs.\ (\ref{sciso1})--(\ref{sc2g}), they can be evaluated easily to 
high numerical precision. After summing them together, they provide a crucial 
check for our calculation of the second-order quasiparticle interaction 
in the partial wave basis (see Section \ref{qp2n}). We set the scalar boson mass
$m_s = 500$ MeV and coupling constant $g_s = 2.5$ and work with
the partial wave matrix elements following from the central potential $V_C(q)$
in eq.\ (\ref{scisop}).

Table \ref{scis} shows the dimensionful Fermi liquid parameters
(labeled `Exact') as obtained from the above analytical formulas 
at nuclear matter
saturation density ($k_F=1.33$ fm$^{-1}$). Due to the simple spin and 
isospin dependence of the underlying interaction the constraint 
$g_L = f_L^\prime = g_L^\prime$ holds. For comparison we show also the
first- and second-order results obtained with the general partial wave expansion. 
The second-order terms are further subdivided into particle-particle, 
hole-hole and particle-hole contributions. We find agreement between both
methods to within 1\% or better for all $L=0,1$ Fermi liquid parameters.
In order to achieve this accuracy, the expansions must be carried out through 
at least the lowest 15 partial waves.

\setlength{\tabcolsep}{.07in}
\begin{table}
\begin{tabular}{|c||c|c|c|c||c|c|c|c|} \hline
\multicolumn{9}{|c|}{Scalar-isoscalar boson exchange ($k_F=1.33$ fm$^{-1}$)} \\ \hline
 & $f_0$ [fm$^2$] & $g_0$ [fm$^2$] & $f^\prime_0$ [fm$^2$] & $g^\prime_0$ [fm$^2$] & 
$f_1$ [fm$^2$] & $g_1$ [fm$^2$] & $f^\prime_1$ [fm$^2$] & $g^\prime_1$ [fm$^2$] \\ \hline 
1st & $-0.809$ & 0.164 & 0.164 & 0.164 & 0.060 & 0.060 & 0.060 & 0.060   \\ \hline \hline
2nd(pp) & $ -0.186$ & 0.056 & 0.056 & 0.056 & 0.038 & $-0.006$ & $-0.006$ & $-0.006$ \\ \hline
2nd(hh) & $-0.033$ & 0.010 & 0.010 & 0.010 & 0.042 & $-0.013$ & $-0.013$ & $-0.013$ \\ \hline
2nd(ph) & 0.198 & 0.061 & 0.061 & 0.061 & 0.100 & 0.085 & 0.085 & 0.085  \\ \hline \hline
Total & $-0.830$ & 0.291 & 0.291 & 0.291 & 0.240 & 0.127 & 0.127 & 0.127  \\ \hline \hline
Exact & $-0.830$ & 0.292 & 0.292 & 0.292 & 0.242 & 0.127 & 0.127 & 0.127  \\ \hline
\end{tabular}
\caption{Fermi liquid parameters ($L=0,1$) for a scalar-isoscalar boson-exchange 
interaction with parameters described in the text. The exact results at $k_F=1.33$ 
fm$^{-1}$ obtained from our derived analytical expressions are compared to the 
numerical results computed via a partial wave expansion.}
\label{scis}
\end{table}

A feature of all realistic NN interactions is the presence of a strong tensor force,
which results in mixing matrix elements between spin-triplet states differing by two
units of orbital angular momentum. At second order these mixing matrix elements 
generate substantial contributions to the $L=0,1$ Fermi liquid parameters. In order to 
test the numerical accuracy of our partial wave expansion scheme for the additional
complexity arising from tensor forces, we consider now the quasiparticle interaction 
in nuclear matter generated by (modified) ``pion'' exchange. To be specific
we take a nucleon-nucleon potential in momentum space of the form
\begin{equation} 
V_T(\vec q\,) = -\frac{g^2}{(m_\pi^2+q^2)^2} \, \vec \sigma_1 
\cdot \vec q\,\,\vec \sigma_2 \cdot \vec q \,\, \vec \tau_1 \cdot \vec \tau_2\, ,
\label{mopem}
\end{equation}
where $g$ is a dimensionless coupling constant and $m_\pi$ a variable ``pion'' mass. 
The isovector spin-spin and tensor potentials in coordinate space following from
$V_T(\vec q\,)$ read
\begin{equation}
\widetilde V_S(r)= \frac{g^2}{24\pi}\frac{e^{-m_\pi r}}{r}(m_\pi r -2)\, , \qquad
\widetilde V_T(r) =\frac{g^2}{24\pi}\frac{e^{-m_\pi r}}{r}(1+ m_\pi r) \, .
\end{equation}
The basic motivation for squaring the propagator in eq.\ (\ref{mopem}) is to 
tame the tensor potential at short distances, and thereby one avoids the linear 
divergence that would otherwise occur in iterated (second-order) one-pion-exchange. 
In the presence of non-convergent loop integrals, analytical and numerical 
treatments become difficult to match properly.  Let us now enumerate the 
contributions at first and second order to the $L=0,1$ Landau parameters
as they arise from modified ``pion'' exchange.

The first-order contributions read
\begin{equation}  {\cal F}^{(1)}_0(k_F) = 
(3-\boldsymbol\sigma)(3-\boldsymbol
\tau){g^2 \over 12m_\pi^2} \bigg[ {1\over 4u^2}\ln(1+4u^2) -{1\over 1+4u^2}
\bigg] \,,  
\label{pm1}
\end{equation}
\begin{equation}  {\cal F}^{(1)}_1(k_F) = 
(3-\boldsymbol\sigma)(3-\boldsymbol
\tau){g^2 \over 4m_\pi^2}\bigg[{1+u^2\over 4u^4}\ln(1+4u^2)-{1\over u^2}+
{1\over 1+4u^2}\bigg] \,,\end{equation}
with the abbreviation $u=k_F/m_\pi$. 
For the second-order contributions we follow the labeling $(a)-(g)$ 
introduced previously for scalar-isoscalar boson exchange:
\begin{equation}  {\cal F}^{(2a)}_0(k_F) = (3-2\boldsymbol\tau){g^4 M_N\over
16 \pi m_\pi^3}\bigg[{3+16u^2 \over 3(1+4u^2)^2}-{5\over 16 u^2}\ln(1+4u^2)
\bigg] \,, 
\end{equation}
\begin{equation}  {\cal F}^{(2a)}_1(k_F) = (3-2\boldsymbol\tau){g^4 
M_N\over 16
\pi m_\pi^3}\bigg[{35\over 8u^2}-{5+24u^2 \over (1+4u^2)^2}-{5\over 32 u^4}
(7+6u^2)\ln(1+4u^2) \bigg] \,, \end{equation}
\begin{eqnarray} {\cal F}^{(2b)}_0(k_F) &=& (5\boldsymbol\tau-3){g^4 M_N 
\over 32
\pi m_\pi^3}\bigg\{ {1+\boldsymbol\sigma \over 8(1+u^2)} +{1\over 3}
(\boldsymbol\sigma-3) \bigg[{1\over 2u^2}\ln{1+2u^2\over1+u^2} \nonumber 
\\ && 
-{1\over 1+2u^2} +\int_0^u\!\! dx\, {1+4x^2+8x^4 \over
u^2(1+2x^2)^3}(\arctan 2x-\arctan x)\bigg] \bigg\} \,,\end{eqnarray}
\begin{eqnarray} {\cal F}^{(2b)}_1(k_F) &=& (5\boldsymbol\tau-3){g^4 M_N 
\over 32
\pi m_\pi^3}\bigg\{ {3\over 4}(1+\boldsymbol\sigma)\bigg[{1\over u^2}-{1 
\over
2(1+u^2)}-{1\over u^4}\ln(1+u^2) \bigg] \nonumber \\ &&  
+(\boldsymbol\sigma-3)
\bigg[{1\over 1+2u^2}-{1\over u^2}+{2+u^2\over 2u^4}\ln{1+2u^2\over 1+u^2}
\nonumber \\ &&  +\int_0^u\!\! dx\, {u^2-2x^2 \over 
u^4(1+2x^2)^3}(1+4x^2+8x^4)
(\arctan 2x-\arctan x)\bigg] \bigg\} \, . \end{eqnarray}
\begin{equation} {\cal F}^{(2c)}_0(k_F)=(3-\boldsymbol\sigma)(3-\boldsymbol\tau)
{4g^4 M_N \over 3\pi^2 m_\pi^3u^2}\int_0^u \!\!dx\,{x^4 \over (1+4x^2)^4} 
\bigg[
2u x +(u^2-x^2) \ln{u+x\over u-x} \bigg] \,, \end{equation}
\begin{equation} {\cal F}^{(2c)}_1(k_F) = 
(3-\boldsymbol\sigma)(3-\boldsymbol
\tau){4g^4 M_N \over \pi^2 m_\pi^3u^4}\int_0^u \!\!dx\,{x^4(u^2-2x^2) \over
(1+4x^2)^4} \bigg[ 2u x +(u^2-x^2) \ln{u+x\over u-x} \bigg] \, .
\end{equation}
\begin{equation} {\cal F}^{(2d)}_0(k_F) =(3-2\boldsymbol\tau) {16g^4 
M_N \over\pi^2
m_\pi^3 u^2}\int_0^u\!\!dx \,{x^5 \over (1+4x^2)^4} \bigg[ u\ln{u+x \over
4(u-x)} +x \ln{u^2-x^2 \over x^2}    \bigg] \,, \end{equation}
\begin{eqnarray} {\cal F}^{(2d)}_1(k_F) &=&(3-2\boldsymbol\tau) 
{16g^4M_N \over
\pi^2 m_\pi^3 u^4}\int_0^u\!\!dx \,{x^5\over (1+4x^2)^4} \bigg[2u^2(u-x)+u^3
\ln{u+x \over 4(u-x)}  \nonumber \\ && \qquad \qquad \qquad \qquad \qquad
\qquad\qquad\quad +x(3u^2-2x^2)\ln{u^2-x^2 \over x^2}\bigg] \,, 
\end{eqnarray}
\begin{equation} {\cal F}^{(2e)}_0(k_F) = (3+2\boldsymbol\tau){16g^4 M_N 
\over
\pi^2 m_\pi^3 u^2}\int_0^u\!\! dx \,{x^5 \over (1+4x^2)^4} \bigg[ 
u\ln{4u^2 \over
u^2-x^2}  -x \ln{u+x \over u-x}    \bigg] \,, \end{equation}
\begin{equation} {\cal F}^{(2e)}_1(k_F) = (3+2\boldsymbol\tau){16g^4 M_N 
\over
\pi^2 m_\pi^3 u^4}\int_0^u\!\! dx \,{x^5\over (1+4x^2)^4} 
\bigg[2u(x^2-u^2)+
u^3\ln{4u^2\over u^2-x^2}  -x^3 \ln{u+x \over u-x}    \bigg] \, .
\end{equation}
We split the crossed terms from the planar-box diagram with Pauli blocking, see Fig.\ 
\ref{qp2norbert}(f),
into factorizable parts:
\begin{equation} {\cal F}^{(2f)}_0(k_F)=(3+\boldsymbol\sigma)(3-5\boldsymbol\tau)
{g^4 M_N\over 96\pi^2 m_\pi^3 u^2} \int_0^u\!\! dx \bigg[{x-u \over 1+(u-x)^2}
-{x+u \over 1+(u+x)^2}+{1\over 2x}\ln{1+(u+x)^2 \over  
1+(u-x)^2}\bigg]^2 \,,
\end{equation}
\begin{equation} {\cal F}^{(2f)}_1(k_F)=(3+\boldsymbol\sigma)(3-5\boldsymbol\tau)
{g^4 M_N \over 96\pi^2 m_\pi^3 u^2}\int_0^u\!\! dx  \Big[G_a^2+2G_b^2\Big]\,,
\end{equation}
\begin{equation} G_a ={u+x \over  1+(u+x)^2}+{x-u \over  
1+(u-x)^2}-{1\over 2u}
\ln{1+(u+x)^2 \over  1+(u-x)^2}\,, \end{equation}
\begin{equation} G_b ={3 \over x}-G_a -{3\over 4u 
x^2}(1+u^2+x^2)\ln{1+(u+x)^2
\over  1+(u-x)^2}\,, \end{equation}
and non-factorizable parts:
\begin{eqnarray} {\cal F}^{(2f')}_0(k_F) 
&=&(\boldsymbol\sigma-3)(3-5\boldsymbol
\tau){2g^4M_N \over 3\pi^2 m_\pi^3 u^2}\int_0^u \!\!dx \,{x^3 \over (1+4x^2)^2}
\int_0^{u-x} \!\!\!dy\,\bigg\{{2u(x-y) (1+4x y) \over R\,[4(x-y)^2+R]} 
\nonumber
\\ && +{1\over R^{3/2}}(u^2-x^2-y^2 +8x^2y^2)\ln{u \sqrt{R} +(1-4x 
y)(x-y) \over
u \sqrt{R} +(4x y-1)(x-y)} \bigg\}  \,, \end{eqnarray}
\begin{eqnarray} {\cal F}^{(2f')}_1(k_F) 
&=&(\boldsymbol\sigma-3)(3-5\boldsymbol
\tau){g^4M_N \over 64\pi^2 m_\pi^3 u^4}\int_0^u \!\!dx 
\,\Bigg\{\bigg[\ln{1+(u+x)^2
\over  1+(u-x)^2}+{1 \over 1+ (u+x)^2}\nonumber \\ && -{1\over  1+(u-x)^2}
\bigg]^2 +{32x^3 \over (1+4x^2)^2} \bigg[ u \ln{4(u-x)\over u+x}-x 
\ln{u^2-x^2
\over x^2} \nonumber \\ && + \int_0^{u-x} \!\!\!dy\,\bigg({4u(x-y)(1+4x y)
(1+2u^2-4x^2)\over R\,[4(x-y)^2+R]} +\Big(2(1+2u^2-4x^2) \nonumber \\ && 
\times
(u^2-x^2-y^2+8x^2y^2) +R\Big) {1\over R^{3/2}}\ln{u \sqrt{R} +(1-4x y)(x-y)
\over u \sqrt{R} +(4x y-1)(x-y)} \bigg) \bigg]\Bigg\} \,, \end{eqnarray}
with auxiliary polynomial $R=4u^2+(4x^2-1)(4y^2-1)$. These two pieces, 
$(2f)$ and $(2f^\prime)$, are distinguished by whether the remaining 
nucleon propagator can be cancelled or not by terms from the product
of (momentum-dependent) $\pi N$ interaction vertices in the numerator.
Finally, the 
density-dependent vertex corrections to modified ``pion'' exchange have nonzero 
crossed terms, which we split again into factorizable parts:
\begin{equation}  {\cal F}^{(2g)}_0(k_F) 
=(3-\boldsymbol\sigma)(3-\boldsymbol
\tau){g^4 M_N \over 96\pi^2 m_\pi^3 u^3 }\bigg[{4u^2\over 
1+4u^2}-\ln(1+4u^2)\bigg]
\bigg[1-{1+2u^2 \over 4u^2}\ln(1+4u^2)\bigg] \,, \end{equation}
\begin{eqnarray}  {\cal F}^{(2g)}_1(k_F) 
&=&(3-\boldsymbol\sigma)(3-\boldsymbol
\tau){g^4 M_N \over 32\pi^2 m_\pi^3 u^5}\bigg[1-{1+2u^2 \over 4u^2}\ln(1+4u^2)
\bigg]\nonumber \\ && \times  \bigg[ 3u^2+{u^2\over 1+4u^2}-(1+u^2) 
\ln(1+4u^2)
\bigg]\,, \end{eqnarray}
and  non-factorizable parts:
\begin{eqnarray} {\cal F}^{(2g')}_0(k_F) &=& 
(3-\boldsymbol\sigma)(3-\boldsymbol
\tau){g^4 M_N \over 24\pi^2 m_\pi^3 u^2} \int_0^u \!\! dx \, \bigg[{4x^2 
\over
1+ 4x^2}-\ln(1+4x^2)\bigg] \nonumber  \\ && \times \bigg\{ {2ux 
(1+4u^2)^{-1}
\over 1+4u^2-4x^2}+{u^2-x^2\over  (1+4u^2-4x^2)^{3/2}} \ln 
{(u\sqrt{1+4u^2-4x^2}
+x )^2 \over(1+4u^2)(u^2-x^2)} \bigg\} \,, \end{eqnarray}
\begin{eqnarray} {\cal F}^{(2g')}_1(k_F) &=& 
(3-\boldsymbol\sigma)(3-\boldsymbol
\tau){g^4 M_N \over 32\pi^2 m_\pi^3 u^4} \int_0^u \!\! dx \, \bigg[{4x^2 
\over
1+ 4x^2}-\ln(1+4x^2)\bigg] \nonumber \\ && \times \bigg\{ {4ux 
(1+2u^2)\over
(1+4u^2)(1+4u^2-4x^2)} -\ln{u+x\over u-x} \nonumber \\ && 
+{1+(u^2-x^2)(6+4u^2)
\over  (1+4u^2-4x^2)^{3/2}} \ln { ( u\sqrt{1+4u^2-4x^2} +x )^2 \over(1+4u^2)
(u^2-x^2)} \bigg\} \, . 
\label{pm2g}
\end{eqnarray}
Together with the coupling constant $g=2.5$ we choose a large ``pion'' mass 
$m_\pi = 400$\,MeV in order to suppress partial wave matrix elements 
from the model interaction $V_T(\vec q\,)$ beyond $J=6$ in the numerical 
computations based on the partial wave expansion scheme. 

We show in Table \ref{mpi} the $L=0,1$ Fermi liquid parameters (at $k_F = 1.33$ fm$^{-1}$) 
for the modified ``pion'' exchange interaction up to second order in perturbation theory. 
The summed results from the analytic formulas eqs.\ (\ref{pm1})--(\ref{pm2g}) are labeled ``Exact'' and 
compared to the results obtained by first evaluating the interaction in the partial
wave basis and then using eqs.\ (\ref{pp2nd})-(\ref{ph2nd}). As in the case of
scalar-isoscalar exchange, we find excellent agreement between the
two (equivalent) methods.

\setlength{\tabcolsep}{.07in}
\begin{table}
\begin{tabular}{|c||c|c|c|c||c|c|c|c|} \hline
\multicolumn{9}{|c|}{Modified ``pion'' exchange ($k_F=1.33$ fm$^{-1}$)} \\ \hline
 & $f_0$ [fm$^2$] & $g_0$ [fm$^2$] & $f^\prime_0$ [fm$^2$] & $g^\prime_0$ [fm$^2$] & 
$f_1$ [fm$^2$] & $g_1$ [fm$^2$] & $f^\prime_1$ [fm$^2$] & $g^\prime_1$ [fm$^2$] \\ \hline 
1st & 0.244 & $-$0.081 & $-$0.081 & 0.027 & $-$0.079 & 0.026 & 0.026 & $-$0.009   \\ \hline \hline
2nd(pp) & $-$0.357 & $-$0.062 & 0.269 & 0.104 & 0.018 & $-$0.005 & 0.027 & 0.009 \\ \hline
2nd(hh) & $-$0.017 & $-$0.002 & 0.009 & 0.003 & 0.029 & 0.003 & $-$0.014 & $-$0.005 \\ \hline
2nd(ph) & 0.146 & $-$0.023 & 0.027 & 0.008 & 0.008 & 0.010 & 0.036 & $-$0.003  \\ \hline \hline
Total & 0.017 & $-$0.169 & 0.224 & 0.142 & $-$0.024 & 0.035 & 0.075 & $-$0.009  \\ \hline \hline
Exact & 0.017 & $-$0.169 & 0.224 & 0.142 & $-$0.023 & 0.035 & 0.074 & $-$0.009  \\ \hline
\end{tabular}
\caption{Fermi liquid parameters ($L=0,1$) for (modified) pion exchange. 
The exact results at $k_F=1.33$ 
fm$^{-1}$ are obtained from our derived analytical expressions and compared to the 
numerical results computed via a partial wave expansion.}
\label{mpi}
\end{table}

\subsection{Realistic nuclear two-body potentials}

After having verified the numerical accuracy of our partial wave expansion scheme,
we extend in this section the discussion to realistic nuclear two-body potentials. 
We start with the Idaho N$^3$LO chiral NN interaction \cite{entem} and employ 
renormalization group
methods \cite{bogner02,bogner03,bogner10} to evolve this (bare) interaction down to a 
resolution scale ($\Lambda \simeq 2$ fm$^{-1}$) at which the NN interaction becomes
universal. The quasiparticle interaction in nuclear matter has been
studied previously with such low-momentum nuclear interactions \cite{schwenk02,holt07,kukei}, 
but a complete second-order calculation has never been performed. Given the
observed better convergence properties of low-momentum interactions in nuclear many-body
calculations, we wish to study here
systematically the order-by-order convergence of the quasiparticle interaction derived 
from low-momentum NN potentials. A complete treatment of low-momentum nuclear forces 
requires the consistent evolution of two- and three-body forces together. We postpone the 
inclusion of contributions to the quasiparticle interaction from the (chiral) three-nucleon 
force to upcoming work \cite{holt11}.

In Table \ref{n3loc} we compare the $L=0,1$ Fermi liquid parameters obtained 
from the bare chiral N$^3$LO potential to those of low-momentum interactions obtained 
by integrating out momenta above a resolution scale of $\Lambda = 2.1$\,fm$^{-1}$ and 
$\Lambda=2.3$\,fm$^{-1}$.
The intermediate-state energies in the second-order diagrams are those of free
nucleons $\epsilon_k = \vec k^2/2M_N$, and we include partial waves up to $J=6$ which result in 
well-converged $L=0,1$ Fermi liquid parameters. Comparing the results at first-order, 
we find a large decrease in the isotropic spin- and isospin-independent Landau 
parameter $f_0$ as the decimation scale decreases. This enhances the (apparent)
instability of nuclear matter against isoscalar density oscillations. The effect results largely 
from integrating out some short-distance repulsion in the bare N$^3$LO interaction. 
A repulsive contact interaction $V_C=4C$ (contributing with equal strength $4C$ in singlet and triplet
$S$-waves) gives rise to a first-order quasiparticle interaction of the form
\begin{equation}
{\cal F}_0^{(1)} = C(3-\vec \sigma_1 \cdot \vec \sigma_2 - \vec \tau_1 \cdot \vec \tau_2
- \vec \sigma_1 \cdot \vec \sigma_2 \, \vec \tau_1 \cdot \vec \tau_2)
\end{equation}
and no contributions for $L\ge 1$. Thus, integrating 
out the short-distance repulsion in the chiral N$^3$LO potential yields a large decrease in 
$f_0$ and a (three-times) weaker increase in $g_0, f_0^\prime$,
and $g_0^\prime$. The increase in $f_0^\prime$ gives rise to an increase in the nuclear 
symmetry energy at saturation density by approximately 20\% for interactions evolved down to 
$\Lambda \simeq 2$ fm$^{-1}$. Overall, the scale dependence of the first-order $L=1$ Landau 
parameters is weaker, and in particular the two isospin-independent ($f_1$ and $g_1$) components of 
the $L=1$ quasiparticle interaction are almost scale independent. However, the parameter
$f_1^\prime$ increases as the cutoff scale is lowered, which results according to eq.\ (\ref{aog}) 
in an increase in the anomalous orbital $g$-factor by 10--15\%.

\setlength{\tabcolsep}{.07in}
\begin{table}
\begin{tabular}{|c||c|c|c|c||c|c|c|c|} \hline
\multicolumn{9}{|c|}{Idaho N$^3$LO potential for $k_F=1.33$ fm$^{-1}$} \\ \hline
 & $f_0$ [fm$^2$] & $g_0$ [fm$^2$] & $f^\prime_0$ [fm$^2$] & $g^\prime_0$ [fm$^2$] & 
$f_1$ [fm$^2$] & $g_1$ [fm$^2$] & $f^\prime_1$ [fm$^2$] & $g^\prime_1$ [fm$^2$] \\ \hline 
1st & $-$1.274 & 0.298 & 0.200 & 0.955 & $-$1.018 & 0.529 & 0.230 & 0.090   \\ \hline \hline
2nd(pp) & $-$1.461 & 0.023 & 0.686 & 0.255 & 0.041 & $-$0.059 & 0.334 & 0.254 \\ \hline
2nd(hh) & $-$0.271 & 0.018 & 0.120 & 0.041 & 0.276 & 0.041 & $-$0.144 & $-$0.009 \\ \hline
2nd(ph) & 1.642 & $-$0.057 & 0.429 & 0.162 & 0.889 & $-$0.143 & 0.130 & 0.142  \\ \hline \hline
Total & $-$1.364 & 0.281 & 1.436 & 1.413 & 0.188 & 0.367 & 0.550 & 0.477 \\ \hline \hline
\multicolumn{9}{|c|}{\vlkn($\Lambda=2.3$ fm$^{-1}$) for $k_F=1.33$ fm$^{-1}$} \\ \hline
 & $f_0$ [fm$^2$] & $g_0$ [fm$^2$] & $f^\prime_0$ [fm$^2$] & $g^\prime_0$ [fm$^2$] & 
$f_1$ [fm$^2$] & $g_1$ [fm$^2$] & $f^\prime_1$ [fm$^2$] & $g^\prime_1$ [fm$^2$] \\ \hline 
1st & $-$1.793 & 0.357 & 0.394 & 1.069 & $-$0.996 & 0.493 & 0.357 & 0.152   \\ \hline \hline
2nd(pp) & $-$0.974 & $-$0.098 & 0.594 & 0.185 & $-$0.129 & $-$0.003 & 0.252 & 0.193 \\ \hline
2nd(hh) & $-$0.358 & 0.030 & 0.169 & 0.075 & 0.338 & 0.028 & $-$0.180 & $-$0.042 \\ \hline
2nd(ph) & 2.102 & 0.095 & 0.588 & 0.254 & 1.512 & 0.003 & 0.204 & 0.329  \\ \hline \hline
Total & $-$1.023 & 0.385 & 1.744 & 1.583 & 0.725 & 0.521 & 0.634 & 0.632  \\ \hline \hline
\multicolumn{9}{|c|}{\vlkn($\Lambda=2.1$ fm$^{-1}$) for $k_F=1.33$ fm$^{-1}$} \\ \hline
 & $f_0$ [fm$^2$] & $g_0$ [fm$^2$] & $f^\prime_0$ [fm$^2$] & $g^\prime_0$ [fm$^2$] & 
$f_1$ [fm$^2$] & $g_1$ [fm$^2$] & $f^\prime_1$ [fm$^2$] & $g^\prime_1$ [fm$^2$] \\ \hline 
1st & $-$1.919 & 0.327 & 0.497 & 1.099 & $-$1.034 & 0.475 & 0.409 & 0.178   \\ \hline \hline
2nd(pp) & $-$0.864 & $-$0.079 & 0.507 & 0.164 & $-$0.130 & 0.011 & 0.236 & 0.174 \\ \hline
2nd(hh) & $-$0.386 & 0.022 & 0.195 & 0.085 & 0.355 & 0.034 & $-$0.195 & $-$0.049 \\ \hline
2nd(ph) & 2.033 & 0.164 & 0.493 & 0.292 & 1.620 & 0.098 & 0.234 & 0.412  \\ \hline \hline
Total & $-$1.135 & 0.434 & 1.692 & 1.640 & 0.812 & 0.617 & 0.684 & 0.715  \\ \hline
\end{tabular}
\caption{Fermi liquid parameters ($L=0,1$) for the Idaho N$^3$LO chiral potential as
well as for the low-momentum nucleon-nucleon interaction \vlk at a cutoff scale of 
$\Lambda=2.1$ and $2.3$ fm$^{-1}$.}
\label{n3loc}
\end{table}

Considering the three parts comprising the second-order quasiparticle interaction, we find
large contributions from both the particle-particle ($pp$) and particle-hole ($ph$) diagrams. 
In particular, the $ph$ term is 
quite large, which suggests the need for an exact treatment of this contribution which 
until now has been absent in the literature. As the decimation scale is lowered, the 
$pp$ contribution is generally reduced while the hole-hole ($hh$) and $ph$ contributions are 
both increased. In previous studies, the $hh$ diagram has often been neglected since it
was assumed to give a relatively small contribution to the quasiparticle interaction. However,
we learn from our exact calculation that its effects are non-negligible for all of the 
spin-independent Landau parameters.

\begin{figure}
\begin{center}
\includegraphics[height=12cm,angle=270]{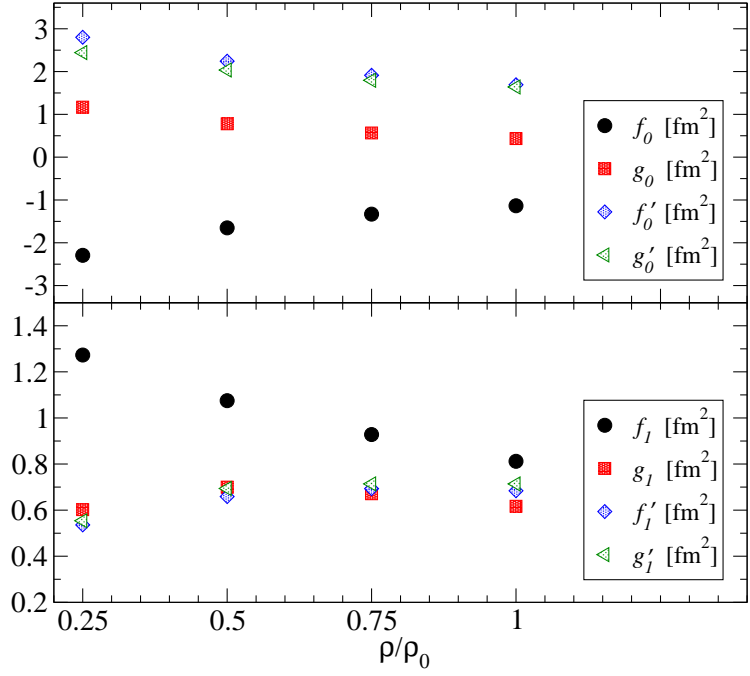}
\end{center}
\vspace{-.5cm}
\caption{Density dependence of the Fermi liquid parameters for the low-momentum
NN interaction $V_{\rm low-k}(\Lambda = 2.1 \, {\rm fm}^{-1})$. Here $\rho_0=0.16$\,fm$^{-3}$ 
is the nuclear matter saturation density.}
\label{dendepflp}
\end{figure}

At second order, the contributions to $f_0$ are sizable and approximately cancel each other 
for the bare N$^3$LO chiral NN interaction,
but they become more strongly repulsive as the resolution scale $\Lambda$ is decreased. This reduces
the large decrease at leading-order in $f_0$ effected through the renormalization group decimation, 
so that after including the second-order corrections, the spread in the values of $f_0$ for all three 
potentials (bare N$^3$LO and its decimations to $\Lambda = 2.1$\,fm$^{-1}$ and $\Lambda = 2.3$\,fm$^{-1}$)
is much smaller than at first order. For each of the three different potentials, the second-order 
terms are strongly coherent in both the $f_0^\prime$ and $g_0^\prime$ channels. In the former
case, this change alone would give rise to a 
dramatic increase the nuclear symmetry energy $\beta$. This effect will be partly reduced through
the increase in the quasiparticle effective mass $M^*$, which we see is now 
close to unity for the bare N$^3$LO chiral interaction but which is strongly scale-dependent 
and enhanced above the free mass $M_N$ as the decimation scale is lowered. The parameter $g_0^\prime$,
related to the energy of Giant Gamow-Teller resonances, is increased by approximately 50\% 
after inclusion of the second-order diagrams. In Fig.\ \ref{dendepflp} we plot the Fermi 
liquid parameters of $V_{\rm low-k}(\Lambda = 2.1 \, {\rm fm}^{-1})$ as a function of density 
$\rho = 2k_F^3/3\pi^2$ from  $\rho_0/4$ to $\rho_0$. We see that all of the $L=0$ parameters, 
together with $f_1$, are enhanced at lower densities.

\subsection{Hartree-Fock single-particle energies}

\begin{figure}
\begin{center}
\includegraphics[height=17cm,angle=270]{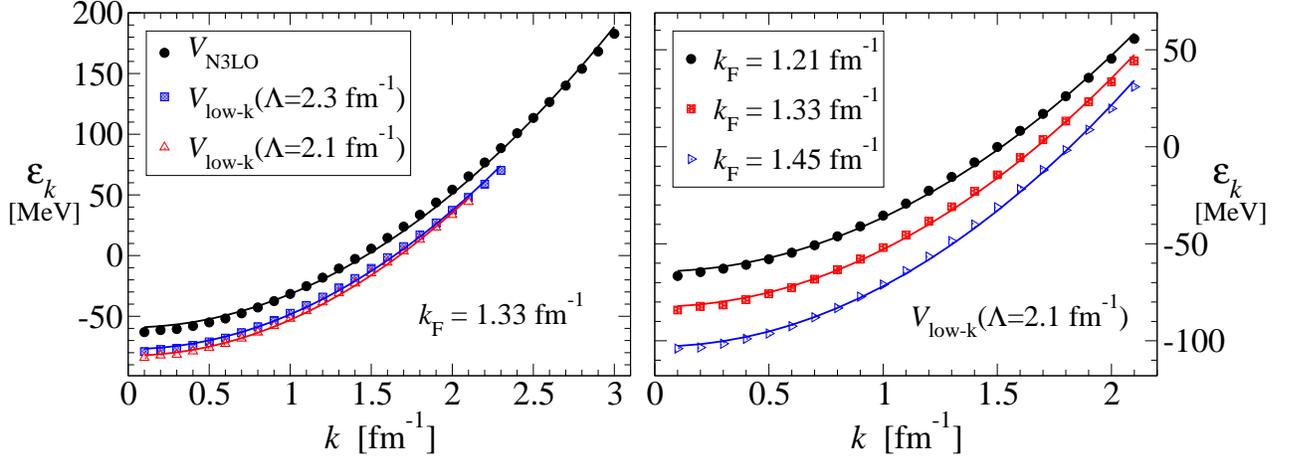}
\end{center}
\vspace{-.5cm}
\caption{Single-particle energies (symbols) computed from eq.\ (\ref{spen}) and fit (lines) 
with the form eq.\ (\ref{disp}) characterized by an effective mass plus energy shift.}
\label{effmassfig}
\end{figure}

\begin{figure}
\begin{center}
\includegraphics[height=14cm,angle=270]{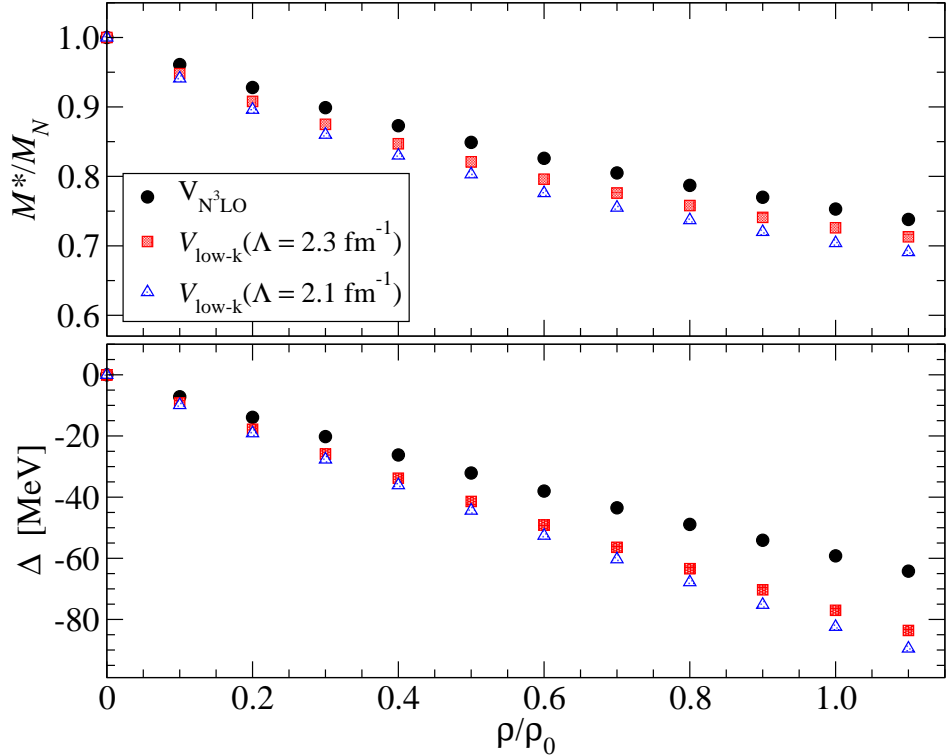}
\end{center}
\vspace{-.5cm}
\caption{Parameterization of the single-particle energy $\epsilon_k = k^2/2M^* + \Delta$ as a 
function of density for three different NN potentials. }
\label{spefig}
\end{figure}

We now discuss the leading-order (Hartree-Fock) contribution to the 
nucleon single-particle energy. The second-order
contributions to the quasiparticle interaction get modified through the resulting change in 
the energy-momentum relation for intermediate-state nucleons. For a nucleon with
momentum $\vec k$, the first-order (in-medium) self-energy correction reads
\begin{eqnarray}
\epsilon_k(k_F) &=& \frac{k^2}{2M_N} + \sum_{s_2, t_2, |\vec k^{\, \prime}| \leq k_F}
\langle \vec k \vec k^{\, \prime} s_1 s_2 t_1 t_2 | \bar V | \ \vec k \vec k^{\, \prime} 
s_1 s_2 t_1 t_2 \rangle \nonumber \\
&=& \frac{k^2}{2M_N} + \frac{1}{2\pi^2}\sum_{lSJT} (2T+1)(2J+1)
\int^{(k+k_F)/2}_{{\rm max} \{0, (k-k_F)/2\} } dp \, p^2\, {\rm min} \{ 
2,(k_F^2-(k-2p)^2)/4pk \} \nonumber \\
&\times& \langle plSJT | \bar V | plSJT \rangle,
\label{spen}
\end{eqnarray}
where $p = |\vec k - \vec k^{\, \prime}\,|/2$. In Fig.\ \ref{effmassfig} 
we plot the single-particle energy as a function of the momentum $k$.  In the left figure 
we show the results for all three NN interactions considered 
in the previous section at a Fermi momentum of $k_F = 1.33$ fm$^{-1}$. In the
right figure we consider only the low-momentum NN interaction with $\Lambda = 2.1$ 
fm$^{-1}$ for three different densities. In all cases one can fit 
the dispersion relation with a parabolic form
\begin{equation}
\epsilon_k = \frac{k^2}{2M^*} + \Delta \, .
\label{disp}
\end{equation}
with $M^*$ the effective mass and $\Delta$ the depth of the single-particle potential.
From the figure one sees that this form holds well across the relevant range of momenta $k$.
In Fig.\ \ref{spefig} we show the extracted effective mass and potential depth
for the three different interactions as a function of the density. 
The energy shift $\Delta$ shows more sensitivity to the decimation scale $\Lambda$ than
the effective mass $M^*$. At saturation density $\rho_0 = 0.16$\,fm$^{-3}$, the variation 
in $\Delta$ is about 30\% while the
spread in $M^*/M_N$ is less than 10\%. Overall, the effective mass extracted from a 
global fit to the momentum dependence of the single-particle energy is in good 
agreement with the local effective mass at the Fermi surface $k=k_F$, encoded in the Landau 
parameter $f_1$. The largest difference occurs for the bare Idaho N$^3$LO potential 
owing to the larger momentum range over which eq.\ (\ref{disp}) is fit to the spectrum.

We employ the quadratic parametrization of the single-particle energy in the second-order
contributions to the quasiparticle interaction eqs.\ 
(\ref{pp2nd})--(\ref{ph2nd}). This greatly simplifies the inclusion of the (first-order) 
in-medium nucleon self energy. The second-order quasiparticle interaction is effectively multiplied
by the same factor $M^*/M_N$, since the constant shift $\Delta$ cancels in the energy 
differences. We then compute the dimensionless Fermi liquid parameters by
factoring out the density of states at the Fermi surface $N_0 = 2M^* k_F/\pi^2$. In Table 
\ref{finaltable} we show the results at $k_F = 1.33$ fm$^{-1}$ for the Idaho N$^3$LO
potential as well as $V_{\rm low-k}^{\Lambda}$ for $\Lambda = 2.1$\,fm$^{-1}$ and 
$\Lambda = 2.3$\,fm$^{-1}$. 
In addition, we have tabulated the theoretical values of the different nuclear
observables that can be obtained from the Fermi liquid parameters.
The quasiparticle effective mass of the bare 
N$^3$LO chiral NN interaction is $M^*/M_N = 0.944$, but this ratio increases beyond 1  
for the low-momentum NN interactions. The inclusion of self-consistent single-particle
energies in the second-order diagrams reduces the very large enhancement in the effective 
mass seen previously in Table \ref{n3loc}. Compared to the other three 
$L=0$ Landau parameters, the spin-spin interaction in nuclear matter is relatively small 
($G_0 = 0.35 - 0.58$). Despite the strong repulsion in $F_0$ that arises from the 
second-order $ph$ diagram, we see that nuclear matter remains unstable against
isoscalar density
fluctuations ($F_0 < -1$), and this behavior is enhanced in evolving the
potential to lower resolution scales. The nuclear symmetry energy $\beta$ is weakly 
scale dependent and we find that the predicted value is within the experimental errors
$\beta = (33 \pm 3)$\,MeV.
Partly due to the rather large effective mass $M^*/M_N$ at the Fermi surface, the anomalous orbital
$g$-factor comes out too small compared to the empirical value of $\delta g_l
=0.23 \pm 0.03$. The $L=0$ spin-isospin Landau parameter $G_0^\prime$ is quite large
for the low-momentum NN interactions. Using the conversion factor $(g_{\pi N}/2M_N)^2 = 1.9$ fm$^2$ 
one gets the values $g_{NN}^\prime = 0.67, 0.75$, 
and 0.77 for the bare N$^3$LO chiral NN interaction and evolved interactions $V_{\rm low-k}^{2.3}$ and 
$V_{\rm low-k}^{2.1}$ respectively. These numbers are in good agreement with values
of $g_{NN}^\prime \agt 0.6$ obtained by fitting properties of giant Gamow-Teller resonances.

The above results highlight the necessity for including three-nucleon force contributions to
the quasiparticle interaction both for the bare and evolved potentials. In fact it has been shown
that supplementing the (low-momentum) potentials
considered in this work with the leading chiral three-nucleon force produces a realistic 
equation of state for cold nuclear matter 
\cite{bogner05,hebeler11}. The large additional repulsion arising in the three-nucleon
Hartree-Fock contribution to the energy per particle should remedy the largest 
deficiency observed in present calculation, namely the large negative value of the
compression modulus ${\cal K}$. A detailed study of the effects of
chiral three-forces (or equivalently the density-dependent
NN interactions derived therefrom \cite{holt09,holt10,hebeler10}) on the Fermi liquid parameters 
is presently underway.

\setlength{\tabcolsep}{.07in}
\begin{table}
\begin{tabular}{|c||c|c|c|c||c|c|c|c||c|c|c|c|} \hline
 & $F_0$ & $G_0$ & $F_0^\prime$ & $G_0^\prime$ & $F_1$ & $G_1$ &
$F^\prime_1$ & $G^\prime_1$ & $M^*/M_N$ & ${\cal K}$ [MeV] & $\beta$ [MeV] & $\delta g_l$ \\ \hline 
$V_{\rm N3LO}$ & $-$1.64 & 0.35 & 1.39 & 1.59 & $-$0.13 & 0.50 & 0.58 & 0.47 
               & 0.96 & $-$148 & 30.5 & 0.12     \\ \hline
$V_{\rm low-k}^{2.3}$ & $-$1.77 & 0.54 & 1.98 & 2.07 & 0.36 & 0.74 & 0.80 & 0.72 
               & 1.12 & $-$152 & 32.5 & 0.07 \\ \hline
$V_{\rm low-k}^{2.1}$ & $-$1.98 & 0.58 & 1.94 & 2.14 & 0.38 & 0.83 & 0.87 & 0.80 
               & 1.13 & $-$191 & 31.8 & 0.07 \\ \hline
\end{tabular}
\caption{Sum of the first- and second-order contributions to the dimensionless Fermi liquid 
parameters for the Idaho N$^3$LO potential and two low-momentum NN interactions 
$V_{\rm low-k}^{\Lambda}$ at $k_F=1.33$ fm$^{-1}$. Hartree-Fock self-energy insertions, as
parameterized in eq.\ (\ref{disp}), are included in the second-order diagrams.}
\label{finaltable}
\end{table}

\section{Summary and conclusions}
We have performed a complete calculation up to second-order for the quasiparticle 
interaction in nuclear matter employing both the Idaho N$^3$LO chiral NN potential as 
well as evolved low-momentum NN interactions. The numerical accuracy of our
results is on the order of 1\% or better. This precision is tested using analytically-solvable 
(at second order) schematic nucleon-nucleon potentials emerging from scalar-isoscalar
boson exchange and modified ``pion'' exchange. We have found that the first-order 
approximation to the full quasiparticle interaction exhibits a strong scale dependence 
in $f_0$, $f_0^\prime$, and $f_1^\prime$, which decreases the nuclear matter 
incompressibility ${\cal K}$ and increases the symmetry energy $\beta$ and anomalous 
orbital $g$-factor $\delta g_l$
as the resolution scale $\Lambda$ is lowered.
Our second-order calculation reveals the importance of the hole-hole contribution in
certain channels as well as the strong effects from the particle-hole contribution 
for the $f_0$ and
$f_1$ Landau parameters. The total second-order contribution has a dramatic effect
on the quasiparticle effective mass $M^*$, the nuclear matter incompressibility ${\cal K}$
and symmetry energy $\beta$, as well as the Landau-Migdal parameter $g_0^\prime$ that
governs the nuclear spin-isospin response. 
In contrast, the components of the spin-spin quasiparticle interaction ($g_0, g_1$) are dominated by the 
first-order contribution. We have included also
the Hartree-Fock contribution to the nucleon single-particle energy, which
reduces the second-order diagrams by about 30\% (as a result of the replacement
$M_N \to M^*$). The final set of $L=0,1$ Landau parameters
representing the quasiparticle interaction in nuclear matter provides a reasonably
good description of the nuclear symmetry energy $\beta$ and 
spin-isospin collective modes. Our calculations demonstrate, however, that the second-order
quasiparticle interaction, generated from realistic {\it two}-body forces only, still
leaves the nuclear many-body system instable with respect to scalar-isoscalar density fluctuations.
Neither the incompressibility of nuclear matter ${\cal K}$ nor the anomalous orbital
$g$-factor $\delta g_l$ could be reproduced satisfactorily (without the inclusion of three-nucleon
forces). A detailed study of the expected improvements in the quasiparticle interaction resulting 
from the leading-order chiral three-nucleon force is the subject of an upcoming 
investigation \cite{holt11}.

\end{document}